\documentclass[preprint,12pt]{elsarticle}
\journal{arXiv, {\rm DRAFT}, \qquad  Corr.~author email: nkuznetsov239@gmail.com}

\usepackage{amsmath,amssymb}
\usepackage{eepic}
\usepackage{hyperref}
\usepackage{graphicx}
\usepackage{enumitem}\biboptions{sort&compress}
\usepackage[left=2.0cm,right=2.0cm,top=1cm,bottom=2cm,bindingoffset=0cm]{geometry}
\usepackage{color}

\newcommand{\be}{\begin{equation}} \newcommand{\ee}{\end{equation}}
\newcommand{\ba}{\begin{aligned}} \newcommand{\ea}{\end{aligned}}

\newtheorem{lemma}{Lemma}

\newtheorem{mycorollary}{Corollary}

\begin{document}

\title{
   Dynamics of the Zeraoulia-Sprott map revisited
}

\author{G. Chen, E.V. Kudryashova, N.V. Kuznetsov, G.A. Leonov}

\address{
City University of Honk Hong, China \\
Saint-Petersburg State University, Russia\\
University of Jyv\"{a}skyl\"{a}, Finland \\
Institute for Problems in Mechanical Engineering of RAS, Russia \\
}

\maketitle

\begin{abstract}
In the paper ``Some Open Problems in
Chaos Theory and Dynamics'' by Zeraoulia and Sprott,
the two-dimensional map
$(x,y) \mapsto (-ax(1+y^2)^{-1},\, x+by)$
was considered and the problem of analytical study
of the boundedness of its attractors was formulated.
In the present paper, the boundedness of its attractors
is studied, the corresponding analytical estimation
of absorbing set is obtained,
and thus an answer to the problem is given.
\end{abstract}



\section{Introduction.}
In the paper \cite{LuWLK-2004},
a 1-D map  related to the dynamics of some  evolutionary processes
(see, e.g. recent surveys on evolutionary algorithms and their
applications to the study of chaotic dynamics
\cite{Zelinka-2015,Zelinka-2016-HA}) was investigated.
Later, its extended two-dimensional version was studied in \cite{ElhadjS-2011-IJBC}:
\begin{equation}\label{sys1}
  \begin{pmatrix}
    x \\ y
  \end{pmatrix}
  \mapsto
  \begin{pmatrix}
    \frac {-ax} {1+y^2}
    \\ x+by
  \end{pmatrix},
\end{equation}
where $a$ and $b$ are bifurcation parameters.
In the paper ``Some Open Problems in
Chaos Theory and Dynamics'' \citep{ElhadjS-2011-OPCSM},
the following problem was formulated:
{\it Find regions in the $a$--$b$ plane in which the map
\eqref{sys1} is bounded and chaotic in the rigorous
mathematical definition of chaos and boundedness
of attractors.}

In the present paper, the boundedness of the attractors of the map (\ref{sys1})
is studied and the corresponding analytical estimation
of absorbing set is obtained, thereby giving an answer to the above problem.

Introduce the notation
\be{}\label{hyk}
   h(y_k) = \frac {-a} {1+ {y_k}^2}, \quad \quad k\in \mathbb{N}_0.
\ee{}
The eigenvalues and eigenvectors of the map are defined by
the following relation:
\[
\begin{pmatrix}
x_{k+1}  \\
y_{k+1}
\end{pmatrix}
=
A
\begin{pmatrix}
x_{k}  \\
y_{k}
\end{pmatrix}
+
\begin{pmatrix}
\frac {ax_{k}{y_{k}}^2} {1+{y_k}^2}  \\
0
\end{pmatrix},
\quad
A= \begin{pmatrix}
-a & 0  \\
1 & b
\end{pmatrix} ,
\]

\[
\det(A-pI)= \det\begin{pmatrix}
-a-p & 0  \\
1 & b-p
\end{pmatrix}
=(-a-p)(b-p)=p^2+p(a-b)-ab.
\]
Thus, the eigenvalues are
\[
 p=b,\quad p=-a.
\]
For the eigenvalue $p=-a$, one has
\[
 \begin{pmatrix}
-a & 0  \\
1 & b
\end{pmatrix}
\begin{pmatrix}
x \\
y
\end{pmatrix}
=
\begin{pmatrix}
-ax  \\
-ay
\end{pmatrix}.
\]
Hence, the corresponding eigenvectors are proportional to
\[
\begin{pmatrix}
  1  \\
  -\frac{1}{a+b}
\end{pmatrix}.
\]

For the eigenvalue $p=b$, one has
\[
 \begin{pmatrix}
  -a & 0  \\
  1 & b
\end{pmatrix}
\begin{pmatrix}
  x \\ y
\end{pmatrix}
=
\begin{pmatrix}
  bx  \\ by
\end{pmatrix},
\]
and, thus, $x=0$.

\bigskip

Next, the following 4 cases are considered, with the associated properties proved:
\\
$|a|<1, |b|<1$:  Global asymptotic stability;\\
$|a|<1, |b|>1$:  Existence of unbounded solutions;\\
$|a|>1, |b|<1$:  Localization of nontrivial global attractor;\\
$|a|>1, |b|>1$:  Existence of unbounded solutions.\\

The cases of $|a|=1$ or $|b|=1$ are not considered
in this paper, which require special consideration
therefore will be discussed elsewhere.

\section{ Case 1: $|a|<1, |b|<1$. Global asymptotic stability.}

\begin{figure}[h!]
\centering
\includegraphics[width=0.95\textwidth]{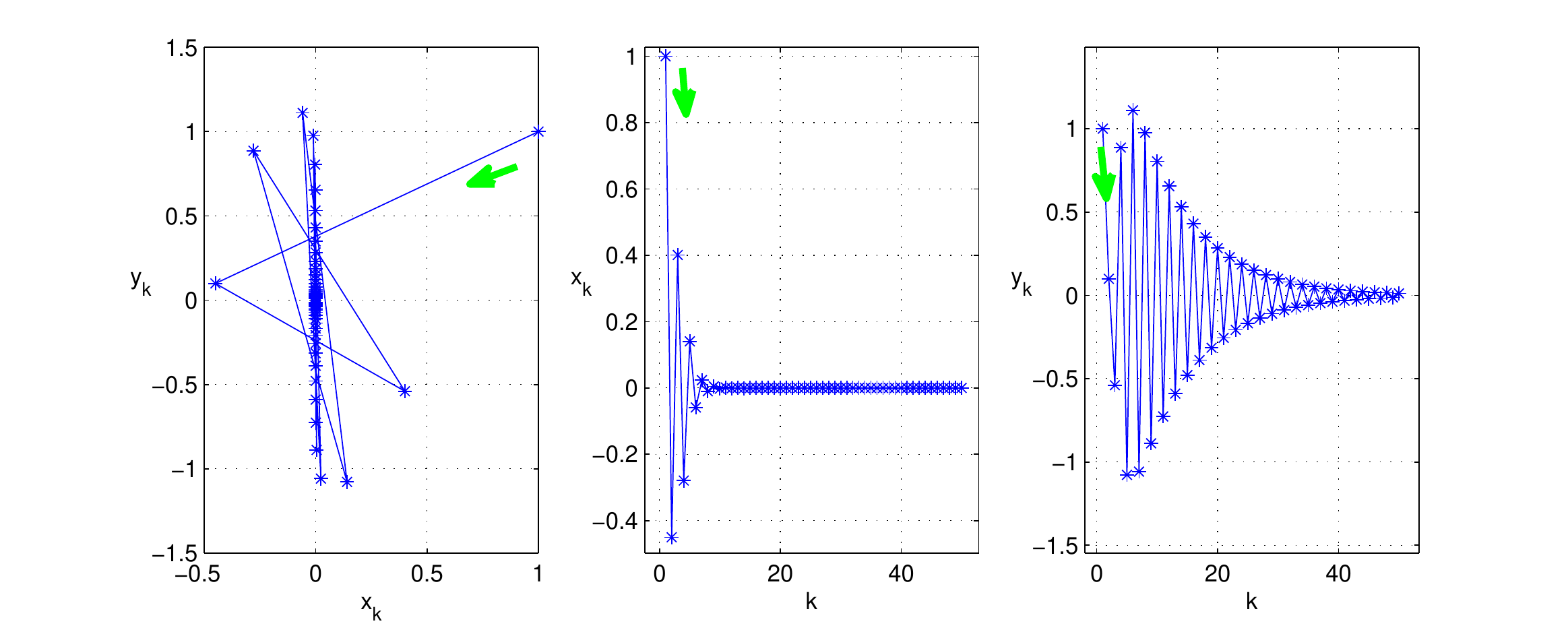}
\caption{Case 1 $(|a|~<~1, |b|~<~1): \quad a~=~0.9,
\quad b~=~-~0.9, \quad x_1~=~1, \quad y_1~=~1, \quad k~=~[1,50]$.}
\label{Case1}
\end{figure}

There exists $\delta>0$ such that $\max(|a|,|b|)<\delta<1$.
Then, $|h(y_k)|<\delta<1$, $\forall k$, and
\[
  |x_k| \le \delta^k |x_0|,\quad \forall k\geq0.
\]
Since $\delta<1$, one has
\[
  x_k \xrightarrow[k\rightarrow \infty]{} 0.
\]
For $y_k$, one has
\[
\begin{aligned}
  |y_{k+1}|  & =  |by_k+x_k| =|b^2y_{k-1}+bx_{k-1}+x_k|
  \\ & \leq |b|^{k+1}|y_{0}|+|b|^k|x_{0}|+|b|^{k-1}|x_{1}|+|b|^{k-2}|x_{2}|+\cdots+|b||x_{k-1}|+|x_k|
  \\ & \leq
  |b|^{k+1}|y_{0}|+ (k+1)\delta^{k}|x_{0}|.
\end{aligned}
\]
Taking into account of the fact that
$|b|<\delta<1$, one gets $y_{k+1} \xrightarrow[k\rightarrow \infty]{} 0$.
Thus, for $|a|<1,|b|<1$, the global asymptotic stability is confirmed
(see Fig.~\ref{Case1}).

\section{ Case 2: $|a|<1, |b|>1$. The existence of unbounded solutions.}
It will be shown that if $|a|<1, |b|>1$, then there exists $(x_0,y_0)$ such that
\[
\begin{aligned}
  &|x_k(x_0,y_0)| \xrightarrow[k\rightarrow \infty]{} 0,\quad
  |y_k(x_0,y_0)|  \xrightarrow[k\rightarrow \infty]{} \infty.
\end{aligned}
\]

\begin{figure}[!h]
  \centering
  \includegraphics[width=0.95\textwidth]{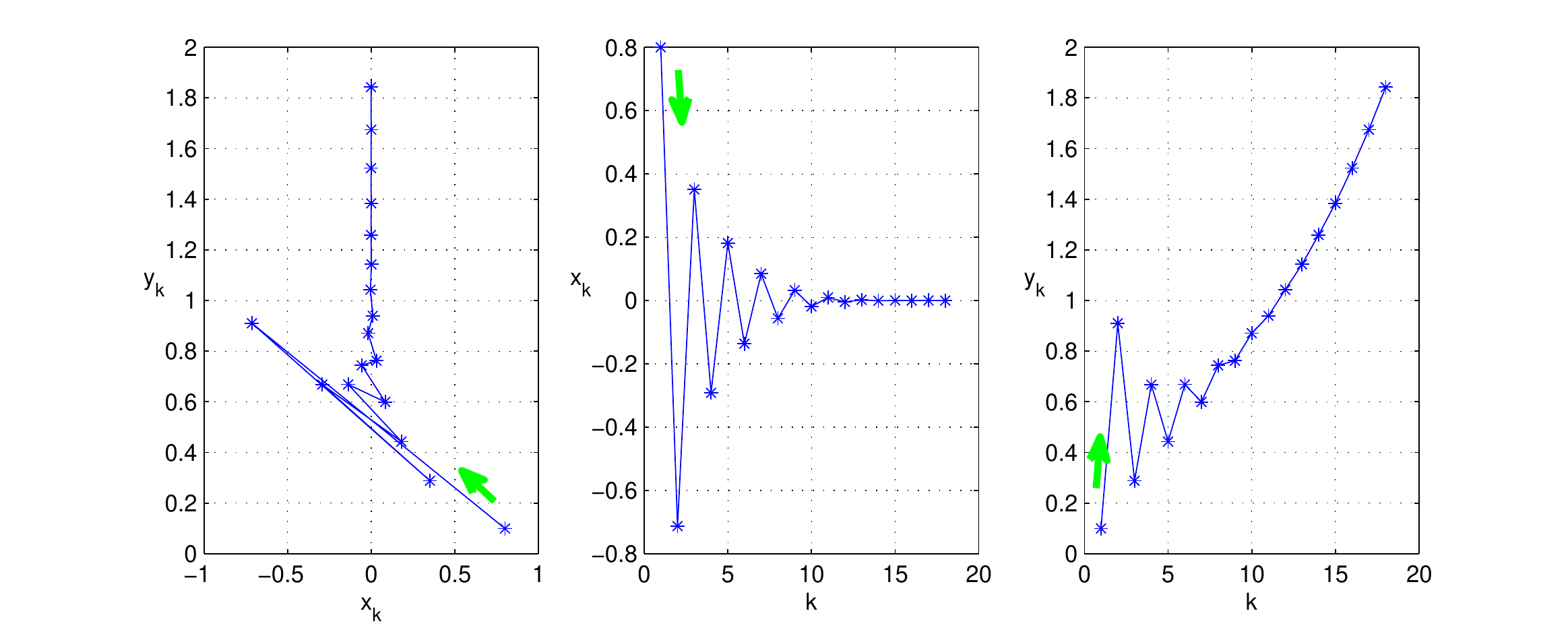}
  \caption{Case 2a $(|a|~<~1, |b|~>~1, b>0):
  \quad a~=~0.9, \quad b~=~1.1, \quad x_1~=~0.8,
  \quad y_1~=~0.1, \quad k~=~[1,18]$.}
  \label{Case2_main}
\end{figure}

\begin{figure}[!h]
  \centering
  \includegraphics[width=1.1\textwidth]{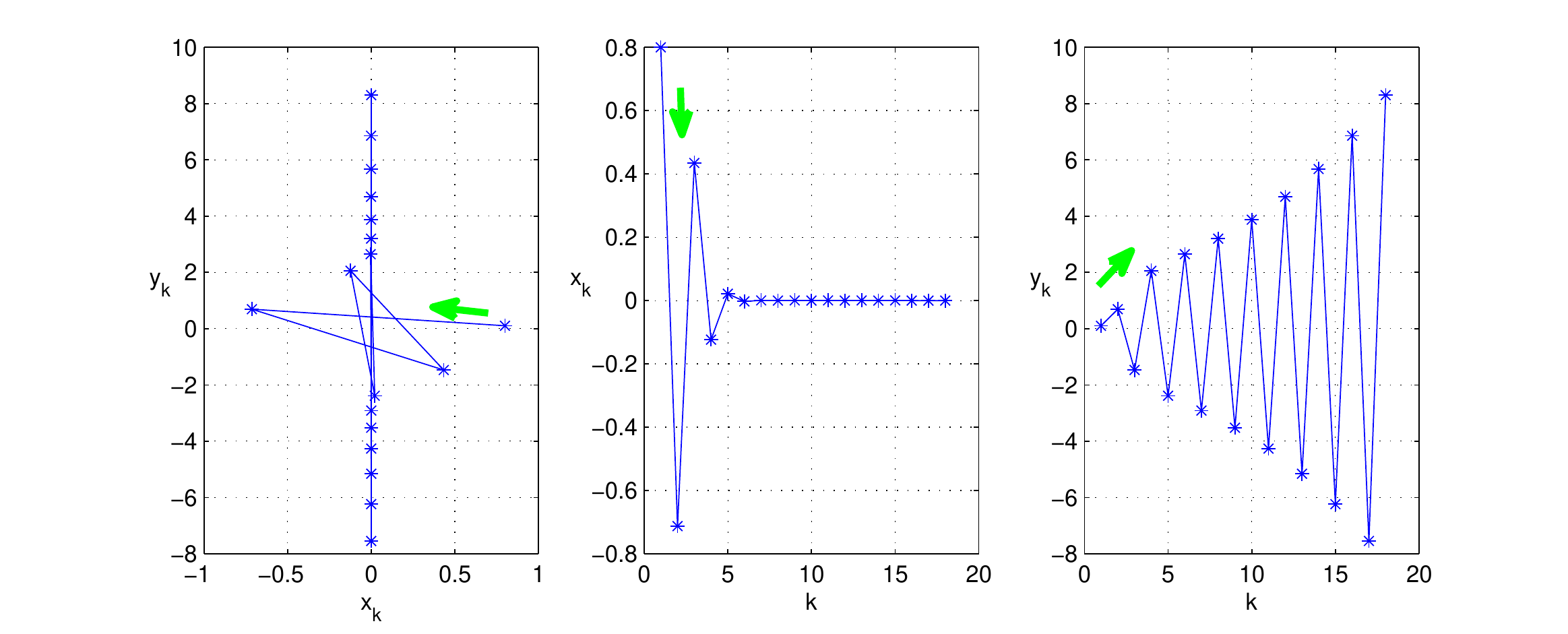}
  \caption{Case 2b $(|a|~<~1, |b|~>~1, b<0):
  \quad a~=~0.9, \quad b~=~-~1.1, \quad x_1~=~0.8,
  \quad y_1~=~0.1, \quad k~=~[1,18]$.}
  \label{Case2_main2}
\end{figure}

For $|a|<1$, there exists $\delta>0$ such that $|a|<\delta<1$.
Then, $|h(y_k)|<\delta<1, \forall k\geq0$, and
\[
  |x_k| \le \delta^k |x_0|,\quad \forall k\geq0.
\]
For $y_k$, one has
\[
\begin{aligned}
  y_{k+1} & = by_k+x_k =b^2y_{k-1}+bx_{k-1}+x_k
  \\ &
  = b^{k+1}y_{0}+b^kx_{0}+b^{k-1}x_{1}+b^{k-2}x_{2}+\cdots+bx_{k-1}+x_k
  \\ &
  = b^{k+1}y_{0}+b^k \left(x_{0}+\frac{x_{1}}{b}+\frac{x_{2}}{b^2}
  +\cdots+\frac{x_{k-1}}{b^{k-1}}+\frac{x_{k}}{b^{k}}\right).
\end{aligned}
\]
Taking into account the fact that $|x_k| \le \delta^k |x_0|$, where $|a|<\delta<1$,
one has
\[
\begin{aligned}
  |y_{k+1}| & \ge |b|^{k+1}|y_{0}|-|b|^k\bigg(|x_{0}|
  +\frac{\delta^1 |x_0|}{|b|}+\frac{\delta^2 |x_0|}{|b|^2}+\cdots
  +\frac{\delta^{k-1} |x_0|}{|b|^{k-1}}+\frac{\delta^{k} |x_0|}{|b|^{k}}\bigg)
  \\ &
  \ge b^{k+1}|y_{0}|-b^k\bigg(|x_{0}|+\frac{\delta^1 |x_0|}{|b|}+\frac{\delta^2 |x_0|}{|b|^2}+\cdots+\frac{\delta^{k-1} |x_0|}{|b|^{k-1}}+\frac{\delta^{k} |x_0|}{|b|^{k}}\bigg).
\end{aligned}
\]
For the sum of the geometric (infinitely decreasing) series,
one has
\[
  S_k\bigg(|x_0|,\frac{\delta}{|b|}\bigg) =
  |x_{0}|+\frac{\delta^1 |x_0|}{|b|}
  +\frac{\delta^2 |x_0|}{|b|^2}+\cdots
  +\frac{\delta^{k-1} |x_0|}{|b|^{k-1}}
  +\frac{\delta^{k} |x_0|}{|b|^{k}}
  \le \frac{|x_0|}{1-\frac{\delta}{|b|}}
  =\frac{|b||x_0|}{|b|-\delta} .
\]
Therefore,
\[
|y_{k+1}| \ge |b|^{k+1}|y_{0}|
-|b|^k\bigg(S_k\big(|x_0|,\frac{\delta}{|b|}\big)\bigg) \ge |b|^{k+1}|y_{0}|-|b|^{k+1}\frac{|x_0|}{|b|-\delta}
\ge |b|^{k+1} \left(|y_0|-\frac{|x_0|}{|b|-1}\right).
\]
If $|y_{0}| \ge \frac{|x_0|}{|b|-1}$, then
$|y_{k+1}(x_0,y_0)|  \xrightarrow[k\rightarrow \infty]{} \infty$
(see Figs.~\ref{Case2_main} and \ref{Case2_main2}).

\section{ Case 3: $|a|>1, |b|<1$. The localization of global attractor.}

Let $0<\delta<1$ and introduce the notations
\[
  R_y = \sqrt{|a|-1+\delta|a|}, \quad  R_x=|b|R_y+\sqrt{a^2-1+\delta a^2}.
\]

\begin{lemma}\label{L1}
If $|a|>1$ and $|b|<1$, then
for any $|x_0|$  and $|y_0| > R_y$, one has
\[
|x_{1}| < |x_0|\frac{1}{1+\delta}.
\]
\end{lemma}
{\bf Proof. }
Since
\[
\begin{aligned}
  |x_0|-|x_{1}| & = |x_0|\bigg(1-\frac{|a|}{1+y_0^2}\bigg)
  \ge |x_0|\bigg(1-\frac{|a|}{1+R_y^2}\bigg)
  \\ &
  = |x_0|\bigg(1-\frac{|a|}{1+|a|-1+\delta|a|}\bigg)
  = |x_0|\bigg(1-\frac{1}{1+\delta}\bigg),
\end{aligned}
\]
one has
\[
  |x_{1}| \le
  |x_0|\frac{1}{1+\delta}.
\]
$\blacksquare$

\begin{lemma} \label{L2}
If $|a|>1$ and $|b|<1$, then
for $|y_0| \le R_y$ and $|x_0| > R_x$, one has
\[
  |x_2| \le |x_0|\frac{1}{(1+\delta)} < |x_0|.
\]
\end{lemma}
{\bf Proof.}
For $y_1$, one has
\[
  |y_1|=|x_0+by_0|\ge |x_0|-|by_0| \geq
  |x_0|-|b|R_y > R_x - |b|R_y = |b|R_y+\sqrt{a^2-1+\delta a^2}
  -|b|R_y = \sqrt{a^2-1+\delta a^2}.
\]
Therefore,
\[
  \frac{a^2}{1+y_1^2} < \frac{a^2}{1+a^2-1+\delta a^2}
  = \frac{1}{1+\delta} <1
\]
and
\[
  |x_0|-|x_2| = |x_0|-\frac{a^2|x_0|}{(1+y_0^2)(1+y_{1}^2)}= |x_0|\bigg(1-\frac{1}{(1+y_0^2)}\frac{a^2}{(1+y_{1}^2)}\bigg) \ge
  |x_0|\bigg(1-\frac{1}{1+\delta}\bigg).
\]
Consequently, $|x_2| \le  |x_0|\frac{1}{1+\delta} < |x_0|$.

\noindent $\blacksquare$

By Lemma~\ref{L1} and Lemma \ref{L2}, one gets the following results.
\begin{mycorollary}\label{Cor1}
For any $x_0, y_0$, there exists $n \in \mathbb{N}_0$ such that
\[
  |x_n| \le R_x.
\]
\end{mycorollary}

\begin{mycorollary}\label{Cor2}
If $|x_0| \le R_x$, then $|x_m| \le a^2R_x$, $\forall m \ge 0$.
\end{mycorollary}
{\bf Proof. }
Let $|x_1| > R_x$.
If $|y_1| > R_y$, then $|x_2| \le |x_1|$.
If $|y_1| \le R_y$, then $|x_3|<|x_1| $.
Therefore,
\[
  |x_m| \le \max(|x_1|,|x_{2}|)\le |ax_{1}|\le a^2|x_{0}|\le a^2R_x, \quad \forall m > 0.
\]
$\blacksquare$

\begin{lemma}\label{L3}
If $|a|>1$ and $|b|<1$, then
for  $|x_0| \le M$ and $|y_0| > \frac{M+\delta}{1-|b|}$,
where $M>0$ and $\delta>0$, one has
\[
 |y_1|<|y_0|-\delta<|y_0|.
\]
\end{lemma}
{\bf Proof. }
Since
\[
\begin{aligned}
  |y_0|-|y_1|=|y_0|-|by_0+x_0|\ge |y_0|-|b||y_0|-|x_0|\ge|y_0|(1-|b|)-|x_0|
  > \frac{M+\delta}{1-|b|}(1-|b|) - M = \delta,
\end{aligned}
\]
one has
\[
 |y_1|<|y_0|-\delta <|y_0|.
\]
$\blacksquare$

\begin{mycorollary}\label{Cor3}
For $|x_0| \le R_x$ and  $|y_0| >
\frac{a^2R_x+\delta}{1-|b|}$, there exists $n \in \mathbb{N}_0$
such that
\[
  |x_n| \le a^2R_x, \quad |y_n| \le \frac{a^2R_x+\delta}{1-|b|}.
\]
\end{mycorollary}
{\bf Proof.}
By Corollary~\ref{Cor2}, one has $|x_n| \le a^2R_x$, $\forall n \ge 0$.
By Lemma \ref{L3}, there exists $n$ such that
$|y_n| \le \frac{a^2R_x+\delta}{1-|b|}$.

$\blacksquare$

\begin{lemma}\label{L4}
If $|a|>1$ and $|b|<1$, then
for  $|x_0| \le M$ and $|y_0| \le \frac{M+\delta}{1-|b|}$, where $M>0$ and $\delta>0$,
one has
\[
 |y_1| < \frac{M+\delta}{1-|b|}.
\]
\end{lemma}
{\bf Proof. }\\
\[
  |y_1|=|by_0+x_0| \le |b||y_0|+|x_0|\le |b|\frac{M+\delta}{1-|b|}+ M
  = \frac{M+|b|\delta}{1-|b|} < \frac{M+\delta}{1-|b|}.
\]
$\blacksquare$\\

\begin{mycorollary}\label{Cor4}
If  $|x_0| \le R_x$, then there exists $N>0$ such that
\[
  |y_n| \le \frac{a^2R_x+\delta}{1-|b|},\quad \forall n>N.
\]
\end{mycorollary}

{\bf Proof. }
By Corollaries~\ref{Cor2} and \ref{Cor3}, there exists $n>0$ such that
\[
  |x_n| \le a^2R_x, \quad |y_n|
  \le \frac{a^2R_x+\delta}{1-|b|},
\]
where $|x_m| \le a^2R_x, \forall m \ge 0$.
It follows from Lemma~\ref{L4} with $M=a^2R_x$ that
\[
  \forall k>n, \quad  |y_k| \le \frac{a^2R_x+\delta}{1-|b|}.
\]
$\blacksquare$

\begin{figure}[h!]
\centering
\includegraphics[width=1.0\textwidth]{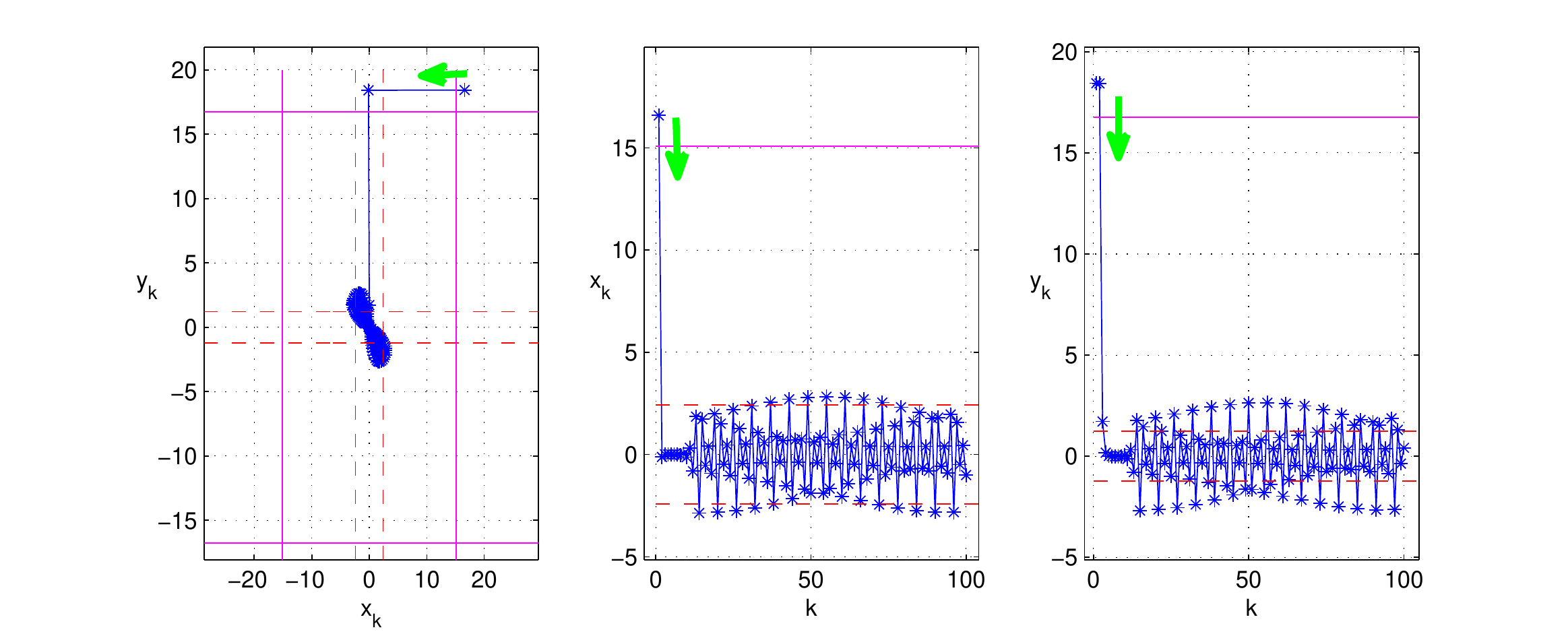}
\caption{Case 3 $(|a|~>~1, |b|~<~1): \quad a~=~2.5,
\quad b~=~0.1, \quad \delta~=~0.00001, \quad x_1~=~16.59,
\quad y_1~=~18.44, \quad k~=~[1,100]$. }
\label{Case3}
\end{figure}

Therefore,
by Corollaries~\ref{Cor1},\ref{Cor2} and \ref{Cor4}
for any $x_0 \ne 0, y_0 \ne 0$, there exists $n>0$
such that, $\forall k>n$,
\[
\begin{aligned}
  & |x_k| \le a^2(|b|\sqrt{|a|-1+|a|\delta}+\sqrt{a^2-1+a^2\delta})=a^2R_x,
  \\ &
  \smallskip
  \\ &
  |y_k| \le \frac{a^2(|b|\sqrt{|a|-1+|a|\delta}
  +\sqrt{a^2-1+a^2\delta})+\delta}{1-|b|}=\frac{a^2R_x+\delta}{1-|b|}.
\end{aligned}
\]
Thus, all possible attractors are placed in the above absorbing set.
Fig.~\ref{Case3} shows the absorbing set and
a self-excited attractor\footnote{
An attractor is called a self-excited attractor if its basin of attraction
intersects with any open neighborhood of an equilibrium;
otherwise, it is called a hidden attractor
\cite{KuznetsovLV-2010-IFAC,LeonovKV-2011-PLA,LeonovKV-2012-PhysD,
LeonovK-2013-IJBC}.
Some recent examples of hidden attractors
in continuous- and discrete-time dynamical systems
can be found in \cite{Chen-2015-IFAC-HA,DancaFKC-2015-IJBC,
Kuznetsov-2016,HeathCS-2015,LeonovKM-2015-EPJST,JafariPGMK-2016-HA}.
}.

\section{ Case 4: $|a|>1, |b|>1$. The existence of unbounded solutions}

Consider a certain $\delta$ satisfying
$|b|^{-1}<\delta<1$.
Let $|y_0| \geq \sqrt{|a|-1}$ and $|x_0|\leq|y_0|(|b|-\delta^{-1})$.
Then,
\[
  |x_1| =\left|\frac{-ax_0}{1+y_0^2}\right| \leq |x_0|, \quad
  |y_1| = |by_0+x_0| \ge |by_0|-|x_0| \geq \delta^{-1}|y_0| > |y_0|,
\]
and  $|x_1|\leq|y_1|(|b|-\delta^{-1})$.
Therefore,
\[
\begin{aligned}
  y_k(y_0) \xrightarrow[k\rightarrow \infty]{} \infty.
\end{aligned}
\]

\begin{figure}[!h]
\centering
\includegraphics[width=1.0\textwidth]{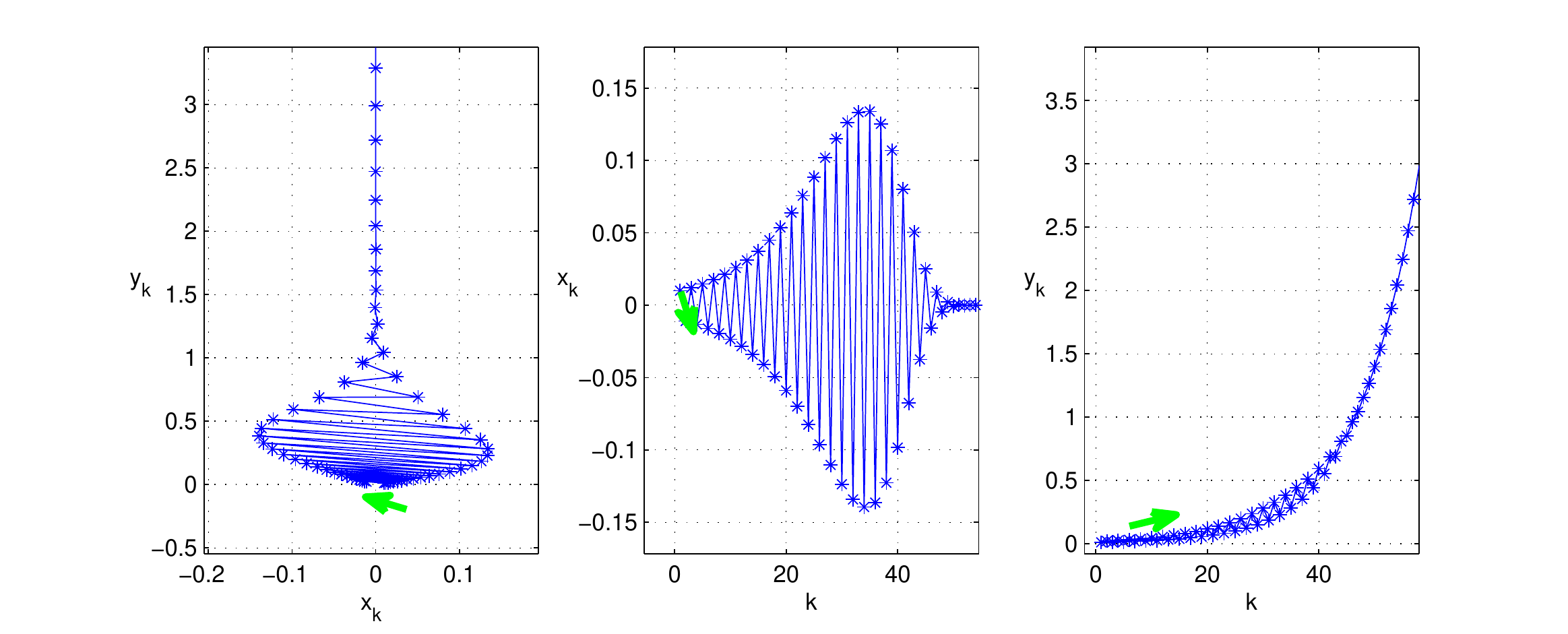}
\caption{Case 4 $(|a|~>~1, |b|~>~1): \quad a~=~1.1,
\quad b~=~1.1, \quad x_1~=~0.01, \quad y_1~=~0.01, \quad k~=~[1,60]$. }
\label{Case4_1}
\end{figure}

Fig.~\ref{Case4_1} 
shows a solution, which tends to infinity.

\section*{Acknowledgment}
This  work  was supported by the Russian Scientific Foundation project 14-21-00041
and the Saint-Petersburg State University.

\bibliographystyle{ws-ijbc}

\end{document}